\documentclass[11pt,a4paper,twoside]{article}  %
\usepackage[american]{babel}     
\usepackage{amsmath,amsthm,amssymb,latexsym}
\usepackage{amsfonts}           
\usepackage{geometry}
\usepackage{graphicx}         
\usepackage{fancyhdr}
\usepackage{tikz}
\usepackage{color}
\usepackage{graphicx}
\usepackage{pgfplots}
\usepackage{amsmath}
\usepackage{amssymb}
\usepackage{caption}
\usepackage{soul}

\def\RR{\mathbb{R}}

\newenvironment { abstract }

\title{ \bf Information Geometric Complexity of a Trivariate Gaussian Statistical Model }

\author{Domenico Felice\\
School of Science and Technology \\University of Camerino, I-62032 Camerino, Italy\\
INFN-Sezione di Perugia, Via A. Pascoli, I-06123 Perugia, Italy
\and
Carlo Cafaro\\
Department of Mathematics\\ Clarkson University, Potsdam, 13699 NY, USA
\and 
Stefano Mancini\\
School of Science and Technology \\University of Camerino, I-62032 Camerino, Italy\\
INFN-Sezione di Perugia, Via A. Pascoli, I-06123 Perugia, Italy
}

\begin{document}

\maketitle

\begin{abstract}
We evaluate the information geometric complexity of entropic motion on
low-dimensional Gaussian statistical manifolds in order to quantify how
difficult it is to make macroscopic predictions about systems in the presence
of limited information. Specifically, we observe that the complexity of such
entropic inferences not only depends on the amount of available pieces of
information but also on the manner in which such pieces are correlated.
Finally, we uncover that, for certain correlational structures, the
impossibility of reaching the most favorable configuration from an entropic
inference viewpoint seems to lead to an information geometric analog of the
well-known frustration effect that occurs in \linebreak statistical physics. 
\end{abstract}

\date{}


\vspace{-12pt}
\section{Introduction}

One of the main efforts in physics is modeling and predicting natural phenomena using relevant information about the system under consideration. Theoretical physics has had a general measure of the uncertainty associated with the behavior of a probabilistic process for more than 100 years: the Shannon entropy \cite{Crutch}. The Shannon information theory was applied to dynamical systems and became successful in describing their unpredictability \cite{Kolmo}.

Along a similar avenue we may set Entropic Dynamics \cite{C02} which makes use of inductive inference (Maximum Entropy Methods \cite{C04}) and Information Geometry \cite{AN00}. This is clearly remarkable given that microscopic dynamics can be far removed from the phenomena of interest, such as in complex biological or ecological systems. Extension of ED to temporally-complex dynamical systems on curved statistical manifolds led to relevant measures of chaoticity \cite{C09}. In particular, an information geometric approach to chaos (IGAC) has been pursued studying chaos in informational geodesic flows describing physical, biological or chemical systems. It is the information geometric analogue of conventional geometrodynamical approaches \cite{P07} where the classical configuration space is being replaced by a statistical manifold with the additional possibility of considering chaotic dynamics arising from non conformally flat metrics.  Within this framework, it seems natural to consider as a complexity measure the (time average) statistical volume explored by geodesic flows, 
namely an Information Geometry Complexity (IGC).

This quantity might help uncover connections between microscopic dynamics and experimentally observable macroscopic dynamics which is a fundamental issue in physics \cite{leb81}. An interesting manifestation of
such a relationship appears in the study of the effects of microscopic
external noise (noise imposed on the microscopic variables of the system) on
the observed collective motion (macroscopic variables) of a globally coupled
map \cite{shib99}. These effects are quantified in terms of the complexity
of the collective motion. Furthermore, it turns out that noise at a
microscopic level reduces the complexity of the macroscopic motion, which in
turn is characterized by the number of effective degrees of freedom of the
system.

The investigation of the macroscopic behavior of complex systems in
terms of the underlying statistical structure of its microscopic degrees of
freedom also reveals effects due to the presence of microcorrelations \cite{ACK10}. In this article we first show which macro-states should be considered in a Gaussian statistical model in order to have a reduction in time of the Information Geometry Complexity. Then, dealing with correlated bivariate and trivariate Gaussian statistical models, the ratio between the IGC in
the presence and in the absence of microcorrelations is explicitly computed, finding an intriguing, even though non yet deep understood, connection with the phenomenon of geometric frustration \cite{SM99}.

The layout of the article is as follows. In Section \ref{sec2} we introduce a general statistical model discussing its geometry and describing both its dynamics and information geometry complexity. In Section \ref{sec3}, Gaussian statistical models (up to a trivariate model) are considered. There, we compute the asymptotic temporal behaviors of their IGCs. Finally, in Section \ref{sec4} we draw our conclusions by outlining our findings and proposing possible further investigations.

\section{Statistical Models and Information Geometry Complexity}\label{sec2}

Given $n$ real-valued random variables $X_1,\ldots, X_n$ defined on the sample space $\Omega$ with joint probability density $p:\RR^n\rightarrow \RR $ satisfying the conditions
\begin{equation}
p(x)\geq 0\; (\forall x\in \RR^n)\quad \mbox{and}\quad \int_{\RR^n} dx \;p(x)=1,
\end{equation}
let us consider a family $\cal P$ of such distributions and suppose that they can be parametrized using $m$ real-valued variables $(\theta^1,\ldots,\theta^m)$ so that
\begin{equation}
{\cal P}=\{p_\theta=p(x|\theta)|\theta=(\theta^1,\ldots,\theta^m)\in\Theta\},
\end{equation}
where $\Theta\subseteq\RR^m$ is the parameter space and the mapping $\theta\rightarrow p_\theta$ is injective. 
In such a way, $\cal P$ is an $m$-dimensional statistical model on $\RR^n$.

The mapping $\varphi:{\cal P}\rightarrow \RR^m$ defined by $\varphi(p_\theta)=\theta$ allows us to consider $\varphi=[\theta^i]$ as a coordinate system for $\cal P$. Assuming parametrizations which are $C^\infty$, we can turn 
$\cal P$ into a $C^\infty$ differentiable manifold (thus, $\cal P$ is called statistical manifold) \cite{AN00}.

The values $x_1,\ldots,x_n$ taken by the random variables define the \emph{micro-state} of the system, while the values  $\theta^1,\ldots,\theta^m$ taken by parameters define the \emph{macro-state} of the system.

Let ${\cal P}=\{p_\theta|\theta\in\Theta\}$ be an $m$-dimensional statistical model. Given a point $\theta$, the Fisher information matrix of $\cal P$ in $\theta$ is the $m\times m$ matrix $G(\theta)=[g_{ij}]$, where the $(i,j)$ entry 
is defined by
\begin{equation}
g_{ij}(\theta):=\int_{\RR^n} dx p(x\arrowvert\theta)\partial_i \log p(x\arrowvert\theta)\partial_j \log p(x\arrowvert\theta),
\label{gFR}
\end{equation}
with $\partial_i$ standing for $\frac{\partial}{\partial\theta^i}$. The matrix $G(\theta)$ is symmetric, positive semidefinite and determines a Riemannian metric on the parameter space $\Theta$ \cite{AN00}. Hence, it is possible to define a Riemannian statistical manifold $\mathcal{M}:=(\Theta,g)$, where   $g= g_{ij}d\theta^i \otimes d\theta^j\ (i,j=1,\ldots,m)$ is the metric whose components  $g_{ij}$ are given by Equation \eqref{gFR} (throughout the paper we use the Einstein sum convention).

Given the Riemannian manifold $\mathcal{M}=(\Theta,g)$, it is well known that there exists only one linear connection $\nabla$(the Levi--Civita connection) on $\mathcal{M}$ that is compatible with the metric $g$ and symmetric~\cite{Lee}. We remark that the manifold $\cal M$ has one chart, being $\Theta$ an open set of $\RR^m$, and the Levi-Civita connection is uniquely defined by means of the Christoffel coefficients 
\begin{equation}
\Gamma_{ij}^k=\frac 1 2 g^{kl}\Big(\frac{\partial g_{lj}}{\partial\theta^i}+\frac{\partial g_{il}}{\partial\theta^j}-\frac{\partial g_{ij}}{\partial\theta^l}\Big),\quad (i,j,k=1,\ldots,m)
\label{Christoff}
\end{equation}
where $g^{kl}$ is the $(k,l)$ entry of the inverse of the Fisher matrix $G(\theta)$.

The idea of curvature is the fundamental tool to understand the geometry of the manifold ${\cal M}=(\Theta,g)$. Actually, it is the basic geometric invariant and the intrinsic way to obtain it is by means of geodesics. It is well-known, that given any point $\theta\in \mathcal{M}$ and any vector $v$ tangent to $\cal M$ at $\theta$, there is a unique geodesic starting at $\theta$ with initial tangent vector $v$. Indeed, within the considered coordinate system, the geodesics are solutions of the following nonlinear second order coupled ordinary differential equations~\cite{Lee}
\begin{equation}\label{geod}
\frac{d^2\theta^k}{d\tau^2}+\Gamma_{ij}^k\frac{d\theta^i}{d\tau}\frac{d\theta^j}{d\tau}=0,
\end{equation}
with $\tau$ denoting the time.

The recipe to compute some curvatures at a point $\theta\in\mathcal{M}$ is the following: first, select a $2$-dimensional subspace $\Pi$ of the tangent space to $\cal M$ at $\theta$; second, follow the geodesics through $\theta$ whose initial tangent vectors lie in $\Pi$ and consider the $2$-dimensional submanifolds $S_{\Pi}$ swiped out by them inheriting a Riemannian metric from $\cal M$; finally, compute the Gaussian curvature of $S_{\Pi}$ at $\theta$, which can be obtained from its Riemannian metric as stated in the \textit{Theorema Egregium} \cite{docarmo}. The number $K(\Pi)$ found in such manner is called the \textit{sectional curvature} of $\cal M$ at $\theta$ associated with the plane $\Pi$. In terms of local coordinates, to compute the sectional curvature we need the curvature tensor,
\begin{equation}
\label{localcurv}
R_{ijk}^h=\frac{\partial\Gamma_{jk}^h}{\partial\theta^i}-\frac{\partial\Gamma_{ik}^h}{\partial\theta^j}+\Gamma_{jk}^l\Gamma_{il}^h-\Gamma_{ik}^l\Gamma_{jl}^h.
\end{equation}
For any basis $(\xi,\eta)$ for a $2$-plane $\Pi\subset T_\theta \mathcal{M}$, the sectional curvature at $\theta\in\mathcal{M}$ is given by \cite{Lee}
\begin{equation}
\label{sectionalcurv}
K(\xi,\eta)=\frac{R(\xi,\eta,\eta,\xi)}{|\xi|^2|\eta|^2-\langle\xi,\eta\rangle},
\end{equation}
where $R$ is the Riemann curvature tensor which is written in coordinates as $R=R_{ijkl}d\theta^i\otimes d\theta^j\otimes d\theta^k\otimes d\theta^l$ with $R_{ijkl}=g_{lh}R^h_{ijk}$ and $\langle\cdot,\cdot\rangle$ is the inner product defined by the metric $g$.

The sectional curvature is directly related to the topology of the manifold; along this direction the \textit{Cartan-Hadamard} Theorem \cite{docarmo} is enlightening by stating that any complete, simply connected $n$-dimensional manifold with non positive sectional curvature is diffeomorphic to $\RR^n$. 

We can consider upon the statistical manifold $\mathcal{M}=(\Theta,g)$ the macro-variables $\theta$ as accessible information and then derive the information dynamical Equation \eqref{geod} from a standard principle of least action of Jacobi type \cite{C02}. The geodesic Equations \eqref{geod} describe a reversible dynamics whose solution is the trajectory between an initial and a final macrostate  $\theta^{ \mbox{\tiny initial}}$ and $\theta^{ \mbox{\tiny final}}$, respectively. The trajectory can be equally traversed in both directions \cite{ACK10}. Actually, an equation relating instability with geometry exists and it makes hope that some global information about the average degree of instability (chaos) of the dynamics is encoded in global properties of the statistical manifolds \cite{P07}. The fact that this might happen is proved by the special case of constant-curvature manifolds, for which the Jacobi-Levi-Civita equation simplifies to \cite{P07}
\begin{equation}\label{JLCE}
\frac{d^2 J^i}{d\tau^2}+K J^i=0,
\end{equation}
where $K$ is the constant sectional curvature  of the manifold (see Equation \eqref{sectionalcurv}) and $J$ is the geodesic deviation vector field. On a positively curved manifold, the norm of the separating vector $J$ does not grow, whereas on a negatively curved manifold, the norm of $J$ grows exponentially in time, and if the manifold is compact, so that its geodesic are sooner or later obliged to fold, this provide an example of chaotic geodesic motion \cite{ca}. 

Taking into consideration these facts, we single out as suitable indicator of dynamical (temporal) complexity,
the information geometric complexity defined as the average dynamical statistical volume~\cite{ca1}
\begin{equation}
\label{IGC}
\widetilde{\textit{vol}}\Big[\mathcal{D}_\Theta^{(\mbox{\tiny geodesic})}(\tau)\Big]:=\frac{1}{\tau}\int_0^\tau d\tau^\prime\textit{vol}\Big[\mathcal{D}_\Theta^{(\mbox{\tiny geodesic})}(\tau^\prime)\Big],
\end{equation}
where
\begin{equation}\label{volIGAC}
\textit{vol}\Big[\mathcal{D}_\Theta^{(\mbox{\tiny geodesic})}(\tau^\prime)\Big]:=\int_{\mathcal{D}_\Theta^{(\mbox{\tiny geodesic})}(\tau^\prime)}\sqrt{\det(G(\theta))}\ d\theta,
\end{equation}
with $G(\theta)$ the information matrix whose components are given by Equation \eqref{gFR}. The integration space $\mathcal{D}_\Theta^{(\mbox{\tiny geodesic})}(\tau^\prime)$ is defined as follows
\begin{equation}\label{parIGAC}
\mathcal{D}_\Theta^{(\mbox{\tiny geodesic})}(\tau^\prime):=\big\{\theta=(\theta^1,\ldots,\theta^m):\theta^k(0)\leq\theta^k\leq\theta^k(\tau^\prime)\big\},
\end{equation}
where $\theta^k\equiv\theta^k(s)$ with $0\leq s\leq\tau^\prime$ such that $\theta^k(s)$ satisfies \eqref{geod}.
The quantity $\textit{vol}\Big[\mathcal{D}_\Theta^{(\mbox{\tiny geodesic})}(\tau^\prime)\Big]$ is the volume of the effective parameter space explored by the system at time $\tau^\prime$. 
The temporal average has been introduced in order to average out the possibly very complex fine details of the entropic dynamical description of the system's complexity dynamics. 

Relevant properties, concerning complexity of geodesic paths
on curved statistical manifolds, of the quantity \eqref{volIGAC} compared to the Jacobi vector field 
are discussed in \cite{CM11}.


\section{The Gaussian Statistical Model}\label{sec3}

In the following we devote our attention to a Gaussian statistical model $\cal P$ whose element are multivariate normal joint distributions for $n$ real-valued variables $X_1,\ldots,X_n$ given by 
\begin{equation}
p(x\arrowvert\theta)=\frac{1}{\sqrt{(2\pi)^n\det C}}\exp\left[-\frac 1 2 (x-\mu)^t {C}^{-1}(x-\mu)\right],
\label{PxT}
\end{equation}
where $\mu=\big(\mathbb{E}(X_1),\ldots,\mathbb{E}(X_n)\big)$ is the $n$-dimensional mean vector and $C$ denotes the $n\times n$ covariance matrix with entries $c_{ij}=\mathbb{E}(X_iX_j)-\mathbb{E}(X_i)\mathbb{E}(X_j)$,
${i,j=1,\ldots,n}$. Since $\mu$ is a $n$-dimensional real vector and $C$ is a $n\times n$ symmetric matrix, the parameters involved in this model should be $n+\frac{n(n+1)}{2}$. Moreover $C$ is a symmetric, positive definite matrix, hence we have the parameter space given by
\begin{eqnarray}
\label{pspace}
\Theta:=\{(\mu,C)|\mu\in\mathbb{R}^n,\ C\in\mathbb{R}^{n\times n}, C>0\}.
\end{eqnarray} 
Hereafter we consider the statistical model given by Equation \eqref{PxT} when the covariance matrix $C$ has only variances $\sigma_i^2=\mathbb{E}(X_i^2)-(\mathbb{E}(X_i))^2$ as parameters. In fact we assume that the non diagonal entry $(i,j)$ of the covariance matrix $C$ equals $\rho\sigma_i\sigma_j$ with $\rho\in\RR$ quantifying the degree of correlation.

We may further notice that the function $f_{ij}(x):=\partial_i\log p(x\arrowvert\theta)\partial_j\log p(x\arrowvert\theta)$, when $p(x\arrowvert\theta)$ is given by Equation~\eqref{PxT}, is a polynomial in the variables $x_i$ ($i=1,\ldots,n$) whose degree is not grater than four. Indeed, we have that 
\begin{equation}
\partial_i\log p(x\arrowvert\theta)=\frac{1}{p(x\arrowvert\theta)}\partial_i p(x\arrowvert\theta)=\partial_i\frac{1}{\sqrt{(2\pi)^n\det C}}+\partial_i \left[-\frac 1 2 (x-\mu)^t {C}^{-1}(x-\mu)\right],
\end{equation}
and, therefore, the differentiation does not affect variables $x_i$. With this in mind, in order to compute the integral in \eqref{gFR}, we can use the following formula \cite{FMP}
\begin{eqnarray}
&&\frac{1}{\sqrt{(2\pi)^n\det C}}\int dx f_{ij}(x) \exp\left[-\frac 1 2 (x-\mu)^t C^{-1} (x-\mu)\right]\nonumber \\
&&=\exp\left[\frac 1 2 \sum_{h,k=1}^n 
c_{hk}\frac{\partial}{\partial x_h}\frac{\partial}{\partial x_k}\right]f_{ij} |_{x=\mu},
\label{Gint}
\end{eqnarray}
where the exponential denotes the power series over its argument (the differential operator).


\subsection{The monovariate Gaussian Statistical Model}\label{mono}

We now start to apply the concepts of the previous section to a Gaussian statistical model of Equation~\eqref{PxT} for $n=1$. In this case, the dimension of the statistical Riemannian manifold $\mathcal{M}=(\Theta,g)$ is at most two. Indeed, to describe elements of the statistical model $\cal P$ given by Equation \eqref{PxT}, we basically need the mean $\mu=\mathbb{E}(X)$ and variance $\sigma^2=\mathbb{E}(X-\mu)^2$. We deal separately with the cases when the monovariate model has only $\mu$ as macro-variable (Case 1), when $\sigma$ is the unique macro-variable (Case 2), and finally when both $\mu$ and $\sigma$ are macro-variables (Case 3).

\subsubsection{Case 1}

Consider the monovariate model  with only $\mu$ as macro-variable by setting $\sigma=1$. In this case the manifold $\cal M$ is trivially the real \textit{flat} straight line, since $\mu\in(-\infty,+\infty)$. Indeed, the integral in \eqref{gFR} is equal to $1$ when the distribution $p(x|\theta)$ reads as $p(x|\mu)=\frac{\exp\big[-\frac{1}{2}(x-\mu)^2\big]}{\sqrt{2 \pi}}$; so the metric is $g=d\mu^2$. Furthermore, from Equations \eqref{Christoff} and \eqref{geod} the information dynamics is described by the geodesic $\mu(\tau)=A_1\tau+A_2$, where $A_1,A_2\in\RR$. Hence, the volume of 
Equation \eqref{volIGAC} results $\textit{vol}\Big[\mathcal{D}_\Theta^{(\mbox{\tiny geodesic})}(\tau^\prime)\Big]=\int d\mu=A_1\tau+A_2$; since this quantity must be positive we assume $A_1,A_2>0$. Finally, the asymptotic behavior of the 
IGC \eqref{IGC} is 
\begin{equation}
\label{IGC1m}
\widetilde{\textit{vol}}\Big[\mathcal{D}_\Theta^{(\mbox{\tiny geodesic})}(\tau)\Big]\approx\Big(\frac{A_1 }{2}\Big)\tau.
\end{equation}
This shows that the complexity linearly increases in time meaning that acquiring information about $\mu$ and updating it, is not enough to increase our knowledge about the micro state of the system.

\subsubsection{Case 2}

Consider now the monovariate Gaussian statistical model of Equation\eqref{PxT} when $\mu=\mathbb{E}(X)=0$ 
and the macro-variable is only $\sigma$. In this case the probability distribution function reads $p(x|\sigma)=\frac{\exp\big[-\frac{x^2}{2\sigma^2}\big]}{\sqrt{2\pi}\sigma}$ while the Fisher--Rao metric becomes $g=\frac{2}{\sigma^2}d\sigma^2$. Emphasizing that also in this case the manifold is flat as well, we derive the information dynamics by means of 
Equations \eqref{Christoff} and \eqref{geod} and we obtain the geodesic $\sigma(\tau)=A_1\exp\big[A_2\tau\big]$. The volume in Equation \eqref{volIGAC} then results 
\begin{equation}
\textit{vol}\Big[\mathcal{D}_\Theta^{(\mbox{\tiny geodesic})}(\tau^\prime)\Big]=\int\frac{\sqrt{2}}{\sigma}d\sigma=\sqrt{2}\log\big[A_1\exp\big[A_2\tau\big]\big].
\end{equation}
Again, to have positive volume we have to assume $A_1,A_2>0$.
Finally, the (asymptotic) IGC \eqref{IGC} becomes
\begin{equation}
\label{IGC1var}
\widetilde{\textit{vol}}\Big[\mathcal{D}_\Theta^{(\mbox{\tiny geodesic})}(\tau)\Big]\approx\Big(\frac{\sqrt{2}A_2}{2}\Big)\tau.
\end{equation}
This shows that also in this case the complexity linearly increases in time meaning that acquiring information about 
$\sigma$ and updating it, is not enough to increase our knowledge about the micro-state of the system.

\subsubsection{Case 3}
The take home message of the previous cases is that we have to account for both mean $\mu$ and variance $\sigma$ as macro-variables to look for possible non increasing complexity.
Hence, consider the probability distribution function is given by,
\begin{equation}
p(x_1,x_2|\mu,\sigma)=\frac{\exp\Big[-\frac{1}{2}\frac{(x-\mu)^2}{\sigma^2}\Big]}{\sigma \sqrt{2\pi}}.
\end{equation}
The dimension of the Riemannian manifold $\mathcal{M}=(\Theta,g)$ is two, where the parameter space $\Theta$ is given by $\Theta=\{(\mu,\sigma)|\mu\in(-\infty,+\infty),\sigma>0\}$ and the Fisher--Rao metric reads as $g=\frac{1}{\sigma^2}d\mu^2+\frac{2}{\sigma^2}d\sigma^2$. Here, the sectional curvature given by Equation \eqref{sectionalcurv} is a negative function and despite the fact that is not constant, we expect a decreasing behavior in time of the IGC. Thanks to Equation \eqref{Christoff}, we find that the only non negative Christoffel coefficients are $\Gamma_{12}^1=-\frac{1}{\sigma}$, $\Gamma_{11}^2=\frac{1}{2\sigma}$ and $\Gamma_{22}^2=-\frac{1}{\sigma}$. Substituting them into Equation \eqref{geod} we derive the following geodesic equations
\begin{equation}\label{geod1}
\left\{\begin{array}{l}
\frac{d^2\mu(\tau)}{d\tau^2}-\frac{2}{\sigma}\frac{d\sigma}{d\tau}\frac{d\mu}{d\tau}=0,\\
\\
\frac{d^2\sigma(\tau)}{d\tau^2}-\frac{1}{\sigma}\Big(\frac{d\sigma}{d\tau}\Big)^2+\frac{1}{2\sigma}\Big(\frac{d\mu}{d\tau}\Big)^2=0.
\end{array}\right.
\end{equation}
The integration of the above coupled differential equations is non-trivial. We follow the method described in \cite{ACK10} and arrive at 
\begin{equation}\label{solmono}
\sigma(\tau)=\frac{2\sigma_0 \exp\Big[\frac{\sigma_0 |A_1|}{\sqrt{2}}\tau\Big]}{1+\exp\Big[\frac{2\sigma_0 |A_1|}{\sqrt{2}}\tau\Big]},\qquad
 \mu(\tau)=-\frac{2\sigma_0\sqrt{2}A_1}{|A_1|\Big(1+\exp\Big[\frac{2\sigma_0 |A_1|}{\sqrt{2}}\tau\Big]\Big)},
\end{equation}
where $\sigma_0$ and $A_1$ are real constants. Then, using \eqref{solmono}, the volume of Equation \eqref{volIGAC} results
\begin{equation}
\textit{vol}\Big[\mathcal{D}_\Theta^{(\mbox{\tiny geodesic})}(\tau^\prime)\Big]=\int\frac{\sqrt{2}}{\sigma^2}d\sigma d\mu=\frac{\sqrt{2}A_1}{|A_1|}\exp\Big[-\frac{\sigma_0 |A_1|}{\sqrt{2}}\tau\Big].
\end{equation}
Since the last quantity must be positive, we assume $A_1>0$. Finally, employing the above expression into Equation \eqref{IGC} we arrive at
\begin{equation}
\label{IGC1}
\widetilde{\textit{vol}}\Big[\mathcal{D}_\Theta^{(\mbox{\tiny geodesic})}(\tau)\Big]\approx\Big(\frac{2}{\sigma_0 A_1}\Big)\frac{1}{\tau}.
\end{equation}
We can now see a reduction in time of the complexity meaning that acquiring information about both $\mu$ and  
$\sigma$ and updating them allows us to increase our knowledge about the micro state of the system.

Hence, comparing Equations \eqref{IGC1m}, \eqref{IGC1var} and \eqref{IGC1} we conclude that the entropic inferences on a Gaussian distributed micro-variable is carried out in a more efficient manner when both its mean and the variance in the form of information constraints are available. Macroscopic predictions when only one of these pieces of information are available are more complex.


\subsection{Bivariate Gaussian Statistical Model}\label{bivar}

Consider now the Gaussian statistical model $\cal P$ of the Equation  \eqref{PxT} when $n=2$. In this case the dimension of the Riemannian manifold $\mathcal{M}=(\Theta,g)$ is at most four.
From the analysis of the monovariate Gaussian model in Section \ref{mono} we have understood that both mean and variance should be considered. Hence the minimal assumption is to consider  $\mathbb{E}(X_1)=\mathbb{E}(X_2)=\mu$ and $\mathbb{E}(X_1-\mu)^2=\mathbb{E}(X_2-\mu)^2=\sigma^2$. 
Furthermore, in this case we have also to take into account the possible presence of (micro) correlations, which appear at the level of macro-states as off-diagonal terms in the covariance matrix.
In short, this implies considering the following probability distribution function
\begin{equation}
p(x_1,x_2|\mu,\sigma)=\frac{\exp\Big[-\frac{1}{2\sigma^2(1-\rho^2)}\Big((x_1-\mu)^2-2\rho(x_1-\mu)(x_2-\mu)+(x_2-\mu)^2\Big)\Big]}{2\pi\sigma^2\sqrt{1-\rho^2}},
\end{equation}
where $\rho\in(-1,1)$.

Thanks to Equation \eqref{Gint} we compute the Fisher-Information matrix $G$ and find
$g=g_{11}d\mu^2 + g_{22}d\sigma^2$ with,
\begin{equation}
\label{gmix}
g_{11}=\frac{2}{\sigma^2(\rho+1)};\ g_{22}=\frac{4}{\sigma^2}.
\end{equation}
The only non trivial Christoffel coefficients \eqref{Christoff} are $\Gamma_{12}^1=-\frac{1}{\sigma}$, $\Gamma_{11}^2=\frac{1}{2\sigma(\rho+1)}$ and $\Gamma_{22}^2=-\frac{1}{\sigma}$. In this case as well, the sectional curvature  
(Equation \eqref{sectionalcurv}) of the manifold $\cal M$ is a negative function and so we may expect a decreasing asymptotic behavior for the IGC. From Equation \eqref{geod} it follows that the geodesic equations are, 
\begin{equation}\label{geod2}
\left\{\begin{array}{l}
\frac{d^2\mu(\tau)}{d\tau^2}-\frac{2}{\sigma}\frac{d\sigma}{d\tau}\frac{d\mu}{d\tau}=0\\
\\
\frac{d^2\sigma(\tau)}{d\tau^2}-\frac{1}{\sigma}\Big(\frac{d\sigma}{d\tau}\Big)^2+\frac{1}{2(1+\rho)\sigma}\Big(\frac{d\mu}{d\tau}\Big)^2=0,
\end{array}\right.
\end{equation}
whose solutions are,
\begin{equation}
\label{solbivar}
\sigma(\tau)=\frac{2\sigma_0 \exp\Big[\frac{\sigma_0 |A_1|}{\sqrt{2(1+\rho)}}\tau\Big]}{1+\exp\Big[\frac{2\sigma_0 |A_1|}{\sqrt{2(1+\rho)}}\tau\Big]},\quad \mu(\tau)=-\frac{2\sigma_0\sqrt{2(1+\rho)}A_1}{|A_1|\Big(1+\exp\Big[\frac{2\sigma_0 |A_1|}{\sqrt{2(1+\rho)}}\tau\Big]\Big)}.
\end{equation}
Using \eqref{solbivar} in Equation \eqref{volIGAC} gives the volume,
\begin{equation}
\label{volbivar}
\textit{vol}\Big[\mathcal{D}_\Theta^{(\mbox{\tiny geodesic})}(\tau^\prime)\Big]=\int\frac{2\sqrt{2}}{\sqrt{1+\rho}\ \sigma^2}d\sigma d\mu=\frac{4A_1}{|A_1|}\exp\Big[-\frac{\sigma_0 |A_1|}{\sqrt{2(1+\rho)}}\tau\Big].
\end{equation}
To have it positive we have to assume $A_1>0$. Finally, employing \eqref{volbivar} in \eqref{IGC} leads to the IGC,
\begin{equation}
\label{IGC2}
\widetilde{\text{vol}}\Big[\mathcal{D}_\Theta^{(\mbox{\tiny geodesic})}(\tau)\Big]\approx\Big(\frac{4\sqrt{2}}{\sigma_0 A_1}\Big)\frac{\sqrt{1+\rho}}{\tau},
\end{equation}
with $\rho \in(-1,1)$. We may compare the asymptotic expression of the ICGs in the presence and in the absence of correlations, obtaining
\begin{equation}
\label{ratio2}
R_{\mbox{\tiny bivariate}}^{\mbox{\tiny strong}}(\rho):=\frac{\widetilde{\textit{vol}}\Big[\mathcal{D}_\Theta^{(\mbox{\tiny geodesic})}(\tau)\Big]}{\widetilde{\textit{vol}}\Big[\mathcal{D}_\Theta^{(\mbox{\tiny geodesic})}(\tau)\Big]_{\rho=0}}=\sqrt{1+\rho},
\end{equation}
where ``strong'' stands for the fully connected lattice underlying the micro-variables.
The ratio $R_{\mbox{\tiny bivariate}}^{\mbox{\tiny strong}}(\rho) $ results a 
\emph{monotonic} increasing function of $\rho $. 

While the temporal behavior of the IGC \eqref{IGC2} is similar to the IGC in \eqref{IGC1}, here correlations play a fundamental role.
From Equation \eqref{ratio2}, we conclude that entropic
inferences on two Gaussian distributed micro-variables on a fully connected
lattice is carried out in a more efficient manner when the two
micro-variables are negatively correlated. Instead, when such micro-variables
are positively correlated, macroscopic predictions become more complex than
in the absence of correlations. 

Intuitively, this is due to the fact that for anticorrelated variables, an increase in one
variable implies a decrease in the other one (different directional change):
variables become more distant, thus more distinguishable in the Fisher--Rao
information metric sense. Similarly, for positively correlated variables, an
increase or decrease in one variable always predicts the same directional
change for the second variable: variables do not become more distant, thus
more distinguishable in the Fisher--Rao information metric sense. This may
lead us to guess that in the presence of anticorrelations, motion on curved
statistical manifolds via the Maximum Entropy updating methods becomes less
complex.


\subsection{Trivariate Gaussian Statistical Model}\label{trivar}

In this section we consider a Gaussian statistical model $\cal P$ of the Equation \eqref{PxT} when $n=3$. In this case as well, in order to understand the asymptotic behavior of the IGC in the presence of correlations between the micro-states, we make the minimal assumption that, given the random vector $X=(X_1,X_2,X_3)$ distributed according to a trivariate Gaussian, then $\mathbb{E}(X_1)=\mathbb{E}(X_2)=\mathbb{E}(X_3)=\mu$ and $\mathbb{E}(X_1-\mu)^2=\mathbb{E}(X_2-\mu)^2=\mathbb{E}(X_2-\mu)^2=\sigma^2$. Therefore, the space of the parameters of $\cal P$ is given by $\Theta=\{(\mu,\sigma)|\mu\in\RR,\sigma>0\}$.

The manifold $\mathcal{M}=(\Theta,g)$ changes its metric structure depending on the number of correlations between micro-variables, namely,
one, two, or three . The covariance matrices corresponding to these cases read, modulo the congruence via a permutation matrix \cite{FMP},
\begin{equation}
C_1=\sigma^2\left(\begin{array}{ccc}
1&\rho&0\\
\rho&1&0\\
0&0&1
\end{array}\right),
\qquad
C_2=\sigma^2\left(\begin{array}{ccc}
1&\rho&\rho\\
\rho&1&0\\
\rho&0&1
\end{array}\right),
\qquad
C_3=\sigma^2\left(\begin{array}{ccc}
1&\rho&\rho\\
\rho&1&\rho\\
\rho&\rho&1
\end{array}\right).
\end{equation}

\subsubsection{Case 1}

First, we consider the trivariate Gaussian statistical model of Equation \eqref{PxT} when $C\equiv C_1$. Then proceeding like in Section \ref{bivar} we have
$g=g_{11}d\mu^2+g_{22}d\sigma^2$, where $g_{11}=\frac{3+\rho}{(1+\rho)\sigma^2}$ and $g_{22}=\frac{6}{\sigma^2}$. Also in this case we find that the sectional curvature of Equation \eqref{sectionalcurv} is a negative function. Hence, as we state in Section \ref{sec2}, we may expect a decreasing (in time) behavior of the information geometry complexity. Furthermore, we obtain the geodesics
\begin{equation}
\sigma(\tau)=\frac{2\sigma_0 \exp\Big[\sigma_0 \sqrt{\mathcal{A}(\rho)}\ \tau\Big]}{1+\exp\Big[2\sigma_0 \sqrt{\mathcal{A}(\rho)}\ \tau\Big]},\ \mu(\tau)=-\frac{2\sigma_0 A_1}{ \sqrt{\mathcal{A}(\rho)}}\frac{1}{1+\exp\Big[2\sigma_0 \sqrt{\mathcal{A}(\rho)}\ \tau\Big]},
\end{equation}
where $\mathcal{A}(\rho)=\frac{A_1^2(3+\rho)}{6(1+\rho)}$ and $A_1\in\RR$. We remark that $\mathcal{A}(\rho)>0$ for all $\rho\in(-1,1)$. Then, the volume~\eqref{volIGAC} becomes
\begin{equation}
\label{voltrivar1}
\textit{vol}\Big[\mathcal{D}_\Theta^{(\mbox{\tiny geodesic})}(\tau^\prime)\Big]=\int \sqrt{\frac{6(3-4\rho)}{(1-2\rho^2)}}\ \frac{1}{\sigma^2}d\sigma d\mu=\frac{6 A_1}{|A_1|}\exp\Big[-\sigma_0 \sqrt{\mathcal{A}(\rho)}\ \tau\Big],
\end{equation}
requiring $A_1>0$ for its positivity.
Finally, using \eqref{voltrivar1} in \eqref{IGC} we arrive at the asymptotic behavior of the IGC 
\begin{equation}
\label{IGC31}
\widetilde{\textit{vol}}\Big[\mathcal{D}_\Theta^{(\mbox{\tiny geodesic})}(\tau)\Big]\approx\Big(\frac{6\sqrt{6}}{\sigma_0 A_1}\Big)\sqrt{\frac{1+\rho}{3+\rho}}\ \frac{1}{\tau}.
\end{equation}
 
Comparing \eqref{IGC31} in the presence and in the absence of correlations yields
\begin{equation}
R_{\mbox{\tiny trivariate}}^{\mbox{\tiny weak}}(\rho):=\frac{\widetilde{\textit{vol}}\Big[\mathcal{D}_\Theta^{(\mbox{\tiny geodesic})}(\tau)\Big]}{\widetilde{\textit{vol}}\Big[\mathcal{D}_\Theta^{(\mbox{\tiny geodesic})}(\tau)\Big]_{\rho=0}}=\sqrt{3}\sqrt{\frac{1+\rho}{3+\rho}},
\label{ratio31}
\end{equation}
where ``weak'' stands for low degree of connection in the lattice underlying the micro-variables

 Notice that $R_{\mbox{\tiny trivariate}}^{\mbox{\tiny weak}}(\rho)$ is a monotonic increasing function of the argument $\rho\in(-1,1)$. 
 
 \subsubsection{Case 2}

When the trivariate Gaussian statistical model of Equation \eqref{PxT} has $C\equiv C_2$, the condition $C>0$ constraints the correlation coefficient to be $\rho\in(-\frac{\sqrt{2}}{2},\frac{\sqrt{2}}{2})$. Proceeding again like in Section \ref{bivar} we have $g=g_{11}d\mu^2+g_{22}d\sigma^2$, where $g_{11}=\frac{3-4\rho}{(1-2\rho^2)\sigma^2}$ and $g_{22}=\frac{6}{\sigma^2}$. The sectional curvature of Equation \eqref{sectionalcurv} is a negative function as well and so we may apply the arguments of Section \ref{sec2} expecting a decreasing in time of the complexity.
Furthermore, we obtain the geodesics
\begin{equation}
\sigma(\tau)=\frac{2\sigma_0 \exp\Big[\sigma_0 \sqrt{\mathcal{A}(\rho)}\ \tau\Big]}{1+\exp\Big[2\sigma_0 \sqrt{\mathcal{A}(\rho)}\ \tau\Big]},\ \mu(\tau)=-\frac{2\sigma_0 A_1}{ \sqrt{\mathcal{A}(\rho)}}\frac{1}{1+\exp\Big[2\sigma_0 \sqrt{\mathcal{A}(\rho)}\ \tau\Big]},
\end{equation}
where $\mathcal{A}(\rho)=\frac{A_1^2(3-4\rho)}{6(1-2\rho^2)}$ and $A_1\in\RR$. We remark that $\mathcal{A}(\rho)>0$ for all $\rho\in(-\frac{\sqrt{2}}{2},\frac{\sqrt{2}}{2})$. Then, the \linebreak volume \eqref{volIGAC} becomes
\begin{equation}
\label{voltrivar2}
\textit{vol}\Big[\mathcal{D}_\Theta^{(\mbox{\tiny geodesic})}(\tau^\prime)\Big]=\int \sqrt{\frac{6(3-4\rho)}{(1-2\rho^2)}}\ \frac{1}{\sigma^2}d\sigma d\mu=\frac{6 A_1}{|A_1|}\exp\Big[-\sigma_0 \sqrt{\mathcal{A}(\rho)}\ \tau\Big].
\end{equation}
We have to set $A_1>0$ for the positivity of the volume \eqref{voltrivar2}, and using it in \eqref{IGC} we arrive at the asymptotic behavior of the IGC
\begin{equation}
\label{IGC32}
\widetilde{\textit{vol}}\Big[\mathcal{D}_\Theta^{(\mbox{\tiny geodesic})}(\tau)\Big]\approx\Big(\frac{6\sqrt{6}}{\sigma_0 A_1}\Big)\sqrt{\frac{1-2\rho^2}{3-4\rho}}\ \frac{1}{\tau}.
\end{equation}

Then, comparing \eqref{IGC32} in the presence and in the absence of correlations yields
\begin{equation}\label{ratio32}
R_{\mbox{\tiny trivariate}}^{\mbox{\tiny mildly weak}}(\rho):=\frac{\widetilde{\textit{vol}}\Big[\mathcal{D}_\Theta^{(\mbox{\tiny geodesic})}(\tau)\Big]}{\widetilde{\textit{vol}}\Big[\mathcal{D}_\Theta^{(\mbox{\tiny geodesic})}(\tau)\Big]_{\rho=0}}=\sqrt{3}\sqrt{\frac{1-2\rho^2}{3-4\rho}},
\end{equation}
where ``mildly weak'' stands for a lattice (underlying micro-variables) neither fully connected nor with minimal connection.

This is a function of the argument $\rho\in(-\frac{\sqrt{2}}{2},\frac{\sqrt{2}}{2})$ that attains the maximum $\sqrt{\frac{3}{2}}$ at $\rho=\frac{1}{2}$, while in the extrema of the interval $(-\frac{\sqrt{2}}{2},\frac{\sqrt{2}}{2})$ it tends to zero.

\subsubsection{Case 3}

Last, we consider the trivariate Gaussian statistical model of the Equation \eqref{PxT} when $C\equiv C_3$. In this case, the condition $C>0$ requires the correlation coefficient to be  $\rho\in(-\frac{1}{2},1)$. 
Proceeding again like in Section \ref{bivar} we have
$g=g_{11}d\mu^2+g_{22}d\sigma^2$, where $g_{11}=\frac{3}{(1+2\rho)\sigma^2}$ and $g_{22}=\frac{6}{\sigma^2}$. We find that the sectional curvature of Equation \eqref{sectionalcurv} is a negative function; hence, we may expect a decreasing (in time) behavior of the complexity. 
It follows the geodesics
\begin{equation}
\label{soltrivar3}
\sigma(\tau)=\frac{2\sigma_0 \exp\Big[\sigma_0 \sqrt{\mathcal{A}(\rho)}\ \tau\Big]}{1+\exp\Big[2\sigma_0 \sqrt{\mathcal{A}(\rho)}\ \tau\Big]},\ \mu(\tau)=-\frac{2\sigma_0 A_1}{ \sqrt{\mathcal{A}(\rho)}}\frac{1}{1+\exp\Big[2\sigma_0 \sqrt{\mathcal{A}(\rho)}\ \tau\Big]},
\end{equation}
where $\mathcal{A}(\rho)=\frac{A_1^2}{2(1+2\rho)}$ and $A_1\in\RR$. We note that $\mathcal{A}(\rho)>0$ for all $\rho\in(-\frac{1}{2},1)$. Using \eqref{soltrivar3}, we compute  
\begin{equation}
\label{voltrivar3}
\textit{vol}\Big[\mathcal{D}_\Theta^{(\mbox{\tiny geodesic})}(\tau^\prime)\Big]=\int \frac{3\sqrt{2}}{\sqrt{(1+2\rho)}}\ \frac{1}{\sigma^2}d\sigma d\mu=\frac{6\sqrt{2} A_1}{|A_1|}\exp\Big[-\sigma_0 \sqrt{\mathcal{A}(\rho)}\ \tau\Big].
\end{equation}
Also in this case we need to assume $A_1>0$ to have positive volume. Finally, substituting Equation \eqref{voltrivar3} into Equation \eqref{IGC}, the asymptotic behavior of the IGC results
\begin{equation}
\label{IGC33}
\widetilde{\textit{vol}}\Big[\mathcal{D}_\Theta^{(\mbox{\tiny geodesic})}(\tau)\Big]\approx\Big(\frac{12}{\sigma_0 A_1}\Big)\sqrt{1+2\rho}\ \frac{1}{\tau}.
\end{equation}

The comparison of \eqref{IGC33} in the presence and in the absence of correlations yields
\begin{equation}\label{ratio3}
R_{\mbox{\tiny trivariate}}^{\mbox{\tiny strong}}(\rho):=\frac{\widetilde{\textit{vol}}\Big[\mathcal{D}_\Theta^{(\mbox{\tiny geodesic})}(\tau)\Big]}{\widetilde{\textit{vol}}\Big[\mathcal{D}_\Theta^{(\mbox{\tiny geodesic})}(\tau)\Big]_{\rho=0}}=\sqrt{1+2\rho},
\end{equation}
where ``strong'' stands for a fully connected lattice underlying the (three) micro-variables.
We remark the latter ratio is a monotonically increasing function of the argument $\rho\in(-\frac{1}{2},1)$. 

\bigskip

\begin{figure} [h]
\centering
\begin{tikzpicture}
\begin{axis}
[xmin=-1,xmax=1,ymin=0,ymax=2,
ytick={1.22},
yticklabels={$\rho_{\mbox{\tiny peak}}$},
xlabel=$\rho$,ylabel=$R(\rho)$]
\addplot [domain=-1:1,
samples=40,smooth,thick,solid]
{sqrt(1+x)};
\addplot [domain=-1:1,
samples=40,smooth,thick,dotted]
{sqrt(3*(1+x)/(3+x))};
\addplot [domain=-0.7071067:0.70710678118654,
samples=40,smooth,thick,dashed]
{sqrt(3*(1-2*x^2)/(3-4*x))};
\addplot [domain=-0.5:1,
samples=40,smooth,thick,densely dashdotted]
{sqrt(1+2*x)};
\end{axis}
\end{tikzpicture}
\caption{Ratio $R(\rho)$ of volumes  \emph{vs.} degree of correlations $\rho$.
Solid line refers to $R_{\mbox{\tiny bivariate}}^{\mbox{\tiny strong}}(\rho)$;
Dotted line refers to $R_{\mbox{\tiny trivariate}}^{\mbox{\tiny weak}}(\rho)$;
Dashed line referes to $R_{\mbox{\tiny trivariate}}^{\mbox{\tiny mildly weak}}(\rho)$;
Dash-dotted refers to $R_{\mbox{\tiny trivariate}}^{\mbox{\tiny strong}}(\rho)$.}
\label{figure}
\end{figure}
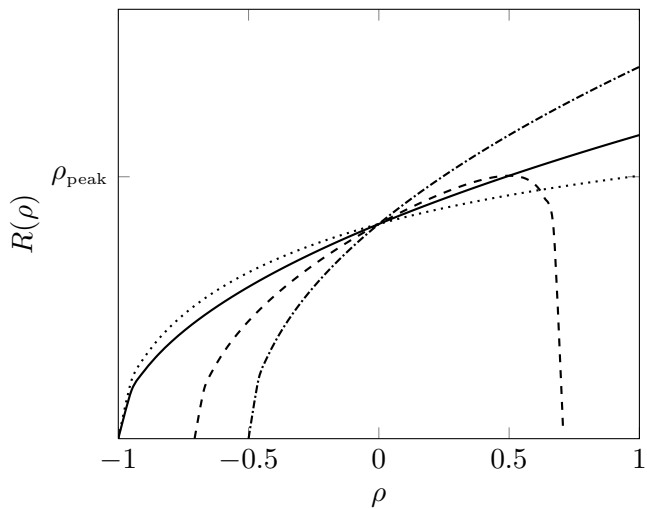

The behaviors of $R(\rho)$ of Equations \eqref{ratio2}, \eqref{ratio31}, \eqref{ratio32} and \eqref{ratio3} are reported in Figure \ref{figure}.

The \emph{non-monotonic} behavior of the
ratio $R_{\mbox{\tiny trivariate}}^{\mbox{\tiny mildly weak}}(\rho)$ in Equation \eqref{ratio32} corresponds to the information geometric complexities for the
mildly weak connected three-dimensional lattice. Interestingly, the growth stops
at a critical value $\rho _{\text{\tiny peak}}=\frac{1}{2}$ at which $
R_{\mbox{\tiny trivariate}}^{\mbox{\tiny mildly weak}}(\rho_{\mbox{\tiny peak}})
=R_{\mbox{\tiny bivariate}}^{\mbox{\tiny strong}}(\rho_{\mbox{\tiny peak}}) $.
From Equation~\eqref{ratio2}, we conclude that entropic inferences on three Gaussian
distributed micro-variables on a fully connected lattice is carried out in a
more efficient manner when the two micro-variables are negatively correlated.
Instead, when such micro-variables are positively correlated, macroscopic
predictions become more complex that in the absence of correlations.
Furthermore, the ratio $R_{\mbox{\tiny trivariate}}^{\mbox{\tiny strong}}(\rho)$ of the information geometric complexities for this fully connected
three-dimensional lattice increases in a \emph{monotonic} fashion. These
conclusions are similar to those presented for the bivariate case. However,
there is a key-feature of the IGC to emphasize when passing from the
two-dimensional to the three-dimensional manifolds associated with fully
connected lattices: the effects of negative-correlations and
positive-correlations are\emph{\ amplified} with respect to the respective
absence of correlations scenarios,
\begin{equation}\label{ratio3su2}
\frac{R_{\mbox{\tiny trivariate}}^{\mbox{\tiny strong}}(\rho)}{
R_{\mbox{\tiny bivariate}}^{\mbox{\tiny strong}}(\rho) }=\sqrt{\frac{
1+2\rho }{1+\rho }}, 
\end{equation}
where $\rho\in(-\frac{1}{2},1)$.

Specifically, carrying out entropic inferences on the higher-dimensional
manifold in the presence of anti-correlations, that is for $\rho \in \left(
-\frac{1}{2},0\right) $, is less complex than on the lower-dimensional
manifold as evident form Equation \eqref{ratio3su2}. The vice-versa is true in the
presence of positive-correlations, that is for $\rho \in \left( 0,
1\right) $.


\section{Concluding Remarks}\label{sec4}

In summary, we considered low dimensional Gaussian statistical models (up to a trivariate model) and have investigated their dynamical (temporal) complexity. This has been quantified by the volume of geodesics for parameters characterizing the probability distribution functions. To the best of our knowledge, there is no \emph{dynamic} measure of complexity
of geodesic paths on curved statistical manifolds that could be compared to
our IGC. However, it could be worthwhile to understand the connection, if any,
between our IGC and the complexity of paths of dynamic systems introduced in
\cite{Bu}. Specifically, according to the Alekseev-Brudno theorem in the algorithmic
theory of dynamical systems \cite{Alek}, a way to predict each new segment of chaotic trajectory is obtained by adding information proportional to the length of this segment and independent of
the full previous length of trajectory. This means that this information cannot be
extracted from observation of the previous motion, even an infinitely long
one! If the instability is a power
law, then the required information per unit time is inversely proportional
to the full previous length of the trajectory and, asymptotically, the
prediction becomes possible. 

For the sake of completeness, we also point
out that the relevance of volumes in quantifying the \emph{static} model
complexity of statistical models was already pointed out in \cite{Bala}
and \cite{Rod}: complexity is related to the volume of a model in
the space of distributions regarded as a Riemannian manifold of distributions
with a natural metric defined by the Fisher--Rao metric tensor. Finally, we
would like to point out that two of the Authors have recently associated
 Gaussian statistical models to networks \cite{FMP}. Specifically, it is
assumed that random variables are located on the vertices of the network while
correlations between random variables are regarded as weighted edges of the
network. Within this framework, a static network complexity measure has been
proposed as the volume of the corresponding statistical manifold. We emphasize
that such a static measure could be, in principle, applied to time-dependent
networks by accommodating time-varying weights on the edges \cite{Mott}.
This requires the consideration of a time-sequence of different statistical
manifolds. Thus, we could follow the time-evolution of a network complexity
through the time evolution of the volumes of the associated manifolds.

 In this work we uncover that in order to have a reduction in time of the complexity one has to consider both mean and variance as macro-variables. This leads to different topological structures of the parameter space in \eqref{pspace}; in particular, we have to consider at least a $2$-dimensional manifold in order to have effects such as a power law decay of the complexity. Hence, the minimal hypothesis in a multivariate Gaussian model consists in considering all mean values equal and all covariances equal. In such a case, however, the complexity shows interesting features depending on the correlation among micro-variables (as summarized in Figure \ref{figure}). 
For a trivariate model with only two correlations the information geometric complexity ratio exhibits a non monotonic behavior in $\rho$ (correlation parameter) taking zero value at the extrema of the range of $\rho$. In contrast to closed configurations (bivariate and trivariate models with all micro-variables correlated each other) the complexity ratio exhibits a monotonic behavior in terms of the correlation parameter.
The fact that in such a case this ratio cannot be zero at the extrema of the range of $\rho$ is reminiscent of the geometric frustration phenomena that occurs in the presence of loops~\cite{SM99}. 

Specifically, recall that a geometrically frustrated system cannot simultaneously minimize
all interactions because of geometric constraints \cite{SM99, moessner}.
For example, geometric frustration can occur in an Ising model which is an
array of spins (for instance, atoms that can take states $\pm 1$) that are
magnetically coupled to each other. If one spin is, say, in the $+1$ state
then it is energetically favorable for its immediate neighbors to be in the
same state in the case of a ferromagnetic model. On the contrary, in
antiferromagnetic systems, nearest neighbor spins want to align in opposite
directions. This rule can be easily satisfied on a square. However, due to
geometrical frustration, it is not possible to satisfy it on a triangle: for
an antiferromagnetic triangular Ising model, any three neighboring spins are
frustrated. Geometric frustration in triangular Ising models can be observed
by considering spin configurations with total spin $J=\pm 1$ and analyzing
the fluctuations in energy of the spin system as a function of temperature.
There is no peak at all in the standard deviation of the energy in the case $%
J=-1$, and a monotonic behavior is recorded. This indicates that the
antiferromagnetic system does not have a phase transition to a state with
long-range order. Instead, in the case $J=+1$, a peak in the energy
fluctuations emerges. This significant change in the behavior of energy
fluctuations as a function of temperature in triangular configurations of
spin systems is a signature of the presence of frustrated interactions in
the system \cite{jay}.

In this article, we observe a significant change in the behavior of the
information geometric complexity ratios as a function of the correlation
coefficient in the trivariate Gaussian statistical models. Specifically, in
the fully connected trivariate case, no peak arises and a monotonic behavior
in $\rho $ of the information geometric complexity ratio is observed. In the mildly
weak connected trivariate case, instead, a peak in the information
geometric complexity ratio is recorded at $\rho _{\text{\tiny peak}}\geq 0$. This
dramatic disparity of behavior can be ascribed to the fact that when
carrying out statistical inferences with positively correlated Gaussian
random variables, the maximum entropy favorable scenario is incompatible
with these working hypothesis. Thus, the system appears frustrated. 

These considerations lead us to conclude that we have uncovered a very
interesting information geometric resemblance of the more standard geometric
frustration effect in Ising spin models. However, for a conclusive
claim of the existence of an information geometric analog of the frustration
effect, we
feel we have to further deepen our understanding. A
forthcoming research project along these lines will be a detailed
investigation of both arbitrary triangular and
square configurations of correlated Gaussian random variables where we take into
consideration both
the presence of different intensities and signs of pairwise interactions ($%
\rho _{ij}\neq \rho _{ik}$ if $j\neq k$, $\forall i$).


\section*{Acknowledgements}

D.F. and S.M. acknowledge the financial support of the Future and Emerging Technologies (FET) programme within the Seventh Framework Programme for Research of the European Commission, under the FET-Open grant agreement TOPDRIM, number FP7-ICT-318121.


\end{document}